\documentclass[a4paper,aps,showpacs,nofootinbib]{revtex4}
\usepackage{graphicx}
\usepackage{amssymb}
\usepackage{bm}
\usepackage{amsmath}

\def\beq{\begin{equation}}
\def\eeq{\end{equation}}

\def\beqn{\begin{eqnarray}}
\def\eeqn{\end{eqnarray}}
\def\nn{\nonumber\\}

\newcommand{\etal}{{\it et al.}}

\begin{document}

\title {Helicity-dependent reaction $\vec \gamma \vec d\to \pi NN$
and its contribution to the GDH sum rule for the deuteron}

\author{M.~I.~Levchuk}\email{levchuk@dragon.bas-net.by}

\affiliation {B.I. Stepanov Institute of Physics, 220072 Minsk,
Belarus}

\date{\today}

\begin{abstract}

Helicity-dependent incoherent pion photoproduction in the reaction
$\vec \gamma \vec d\to \pi NN$
is studied in the framework of the diagrammatic approach.
Contributions to the reaction amplitude from diagrams corresponding
to impulse approximation as well as $NN$ and $\pi N$
interactions in the final state have been evaluated.
The elementary $\gamma N\to\pi N$ operator is taken from the
MAID and SAID models.
A detailed comparison of the predictions with recent experimental data by
the GDH and A2 collaborations at energies below 500 MeV is presented.
Reasonable agreement with the data on the yields and cross sections
for polarized beam and polarized target has been achieved in
all channels. The unpolarized data of the GDH and A2 collaborations
have also been analyzed within the approach.
A strong overestimation for the neutral channel has been found.
At the same time, the model provides a quite satisfactory
description of the unpolarized data for the charged channels.
The sensitivity of the obtained results to the choice
of the elementary $\gamma N\to\pi N$ operator is discussed in detail.
The contribution of the $\gamma d\to\pi NN$ reaction to the GDH sum rule
for the deuteron up to a photon energy of 1.65~GeV has been evaluated
with the result of $235\pm 25$~$\mu$b.

\end{abstract}


\maketitle

\section{Introduction}
\label{intr}

For more than a decade, the GDH and A2 collaborations have been carrying out
intense experimental studies aimed at the verification of the famous
Gerasimov-Drell-Hearn (GDH) sum rule~\cite{gerasimov65,drell-hearn65}.
A comprehensive overview concerning the status of these experiments
is given in Ref.~\cite{helbing06}.

The GDH sum rule gives an integral relation between anomalous
magnetic moment $\kappa$ of a spin-$S$ particle and the difference
of the total photoabsorption cross sections with parallel,
$\sigma_P$, and antiparallel, $\sigma_A$, photon and target spin
alignments and reads
\beq
\label{GDH}
\frac
{4\pi^2\alpha\kappa^2}{m^2}~S= \int^\infty_0 \frac
{\sigma_P(E_\gamma)-\sigma_A(E_\gamma)}{E_\gamma}~dE_\gamma,
\eeq
where $E_\gamma$  is the photon laboratory energy, $m$ is the
particle mass, and $\alpha=1/137$. Equation (\ref{GDH}) shows that
ground-state properties given by $\kappa$ and $m$ are related to
the energy-weighted excitation spectrum of the particle. The GDH
sum rule relies on the basic physics principles of Lorenz and gauge
invariance, unitarity, crossing symmetry, and an unsubtracted
dispersion relation applied to the forward Compton amplitude.
Therefore, measurements of the right-hand side (rhs) of this
equation can serve as a fundamental cross-check of these
principles.

The left-hand side (lhs) of Eq.~(\ref{GDH}) for the proton
is $205$~$\mu$b. The integrand in the rhs
of Eq.~(\ref{GDH}) at energies from 0.2 to 2.9 GeV was determined
by the GDH and A2 collaborations
in a series of experiments with circularly polarized photons
and longitudinally polarized protons~\cite{ahrens01,dutz03,dutz04}.
Involving theoretical predictions from threshold to 0.2 GeV
and above 2.9~GeV, the GDH group obtained
for the rhs of Eq.~(\ref{GDH})
$212\pm 6({\rm stat})\pm 16 ({\rm syst})$~$\mu$b, which
is in good agreement with the GDH sum rule prediction.

The check of the isospin structure of the GDH sum rule
requires measurements on the neutron.
However, because of the absence of a stable, dense, free-neutron
target, direct measurements for the neutron are impossible
so that information on the neutron can be extracted from
deuteron data. Experiments with circularly polarized photons
and longitudinally polarized deuterons were also performed
by the GDH and A2
collaborations~\cite{krusche99,dutz05,ahrens06,ahrens09,ahrens10}
as well as by the LEGS Collaboration~\cite{LEGS}.
In fact, however, a contribution to the neutron GDH integral
from deuteron data was extracted only
in Ref.~\cite{dutz05} and therewith in the limited energy range
of 815--1825 MeV. The extraction procedure
relied mainly on an assumption that at these energies
the incoherent, quasi-free meson production reactions
dominate; i.e. if one supposes that the equality
$\Delta \sigma_d=\Delta \sigma_p+\Delta \sigma_n$
($\Delta \sigma=\sigma_P-\sigma_A$) is valid.
Such an assumption seems to be in general agreement
with the results obtained by
Arenh\"ovel, Fix, and Schwamb (AFS)~\cite{AFS04}.
Using theoretical predictions for the unmeasured energy regions, the
authors of Ref.~\cite{dutz05} obtained the value of
$226$~$\mu$b for the rhs of Eq.~(\ref{GDH}) which
is in agreement with the GDH sum rule value of $233$~$\mu$b.
Of course, this agreement should not be considered
as an experimental verification of the GDH sum rule
for the neutron, because the net experimental contribution
to this value is only about $15\%$.

The helicity-dependent cross section on the deuteron
in the energy range $200<E_\gamma<800$~MeV was also measured
in Ref.~\cite{ahrens09}.
Combining the obtained data with those from Ref.~\cite{dutz05}
and using again the assumption on the incoherency of the
proton and neutron contributions in the range
$200<E_\gamma<1800$~MeV, the authors of Ref.~\cite{ahrens09}
obtained the value of $197$~$\mu$b in this range.
The said assumption seems to be rather rough
below, say, 500 MeV, where the $\gamma d\to np$ and
the $\gamma d\to\pi^0d$ reactions contribute to the
polarized total cross section, nevertheless,
the extracted value indicates that
the GDH integrals for the proton and neutron
are of the same order of magnitude.

Together with the verification of the GDH sum rule for the
neutron, the helicity-dependent cross sections on the deuteron
are also required to test this sum rule for the deuteron itself.
The anomalous magnetic moment of the deuteron is very small
($\kappa_d=-0.143$~nm) because, first, the anomalous magnetic
moments of the proton and neutron are almost equal in
magnitude but opposite in sign and, second, nucleon spins
in the deuteron are predominantly parallel. In view of this,
the lhs of Eq.~(\ref{GDH}) for the deuteron is $0.65$~$\mu$b,
which is more than two orders smaller than the nucleon values.
Therefore, there should be
large negative contributions almost totally canceling
the nucleon contribution. This point was discussed for the
first time in Ref.~\cite{arenh97}. It was found that a huge
negative contribution amounting to about  $-400$~$\mu$b
indeed exists and stems from the deuteron photodisintegration
reaction.
Later on, a comprehensive analysis of different
channels in the deuteron GDH integral was
performed within the AFS framework~\cite{AFS04}. They
evaluated effects from deuteron photodisintegration and
single and double-pion production as well as from single-$\eta$
production with the total result of $27.31$~$\mu$b.

If one intends to extract the neutron value
for the GDH integral from deuteron data
or verify the deuteron GDH sum rule,
in either case one has to be confident that theoretical
models used in analyses of these data provide
reliable descriptions of the helicity-dependent cross sections
for different reactions on the deuteron.
As was found in Refs.~\cite{arenh97,AFS04,darwish07},
a large contribution to the deuteron GDH integral stems from
incoherent single-pion production $\gamma d\to\pi NN$.

Intense experimental investigations of the reaction
$\vec\gamma \vec d\to\pi NN$
have recently been performed by the GDH and A2
collaborations~\cite{ahrens06,ahrens09,ahrens10}.
References~\cite{ahrens06,ahrens09} have provided data on
the helicity-dependent total inclusive and semiexclusive
cross sections in the energy range from 200 to 800~MeV.
In the very recent work of the same collaborations~\cite{ahrens10},
the helicity dependence of the differential cross section in
the $\vec\gamma \vec d\to\pi NN$ channel has been measured for the first
time in the $\Delta$ region. There are also measurements
of the helicity-dependent total cross section of the reaction
$\vec\gamma \vec d\to\pi^0X$~\cite{krusche99,LEGS}, where $X=np$
(the incoherent channel) or $X=d$ (the coherent channel).

The main goal of the present work is to analyze the data
of Refs.~\cite{ahrens06,ahrens09,ahrens10} on the
$\vec\gamma \vec d\to\pi NN$ reaction in the framework of
the diagrammatic model built in Ref.~\cite{ls06}. In that work,
we restricted ourselves to the discussion of unpolarized and
single spin-dependent observables due to the lack of data on double
spin-dependent observables at the time when the work was being prepared.
It was found that the model~\cite{ls06} provided a quite
satisfactory description of all available data except for
the unpolarized differential and total cross
sections in the $\gamma d\to\pi^0np$ channel for which the data
were notably overestimated.
In the present work, we compare predictions
of the approach~\cite{ls06} to the polarized data from
Refs.~\cite{ahrens06,ahrens09,ahrens10} with special
emphasis on the model dependence of our results.
Sources for theoretical uncertainties are discussed
in Ref.~\cite{ls06}, to which the interested reader is referred.

Another principal goal of this work is to evaluate the contribution
from the $\gamma d\to\pi NN$ channel to the deuteron GDH integral.
This task requires an extension of the model up to about 1.5--2~GeV.
Although in Ref.~\cite{ls06} we restricted our consideration  to
the $\Delta$ resonance region,  the model is applicable
at higher energies, too. As was shown in Ref.~\cite{FixAr05},
one can expect that in the framework of the diagrammatic approach,
contributions due to $NN$ and $\pi N$ final state interaction (FSI)
are reliably calculated up to a photon energy of 800~MeV.
Taking into account that these contributions are small
above 500 MeV, they can obviously  be disregarded
for $E_\gamma\agt 800$~MeV.
Hence, the upper limit for the applicability of the model
depends on that for the impulse approximation (IA) which, in turn,
depends on the applicability domain of an employed elementary operator
$\gamma N\to\pi N$ of pion production on the nucleon.
In the framework of the approach~\cite{ls06},
this operator is taken from multipole analyses
SAID~\cite{SAID,Dugger07,Dugger09} and
MAID~\cite{MAID} and can be used up to 2~GeV for the first analysis
and up to 1.65~GeV for the second one.

The contribution from the $\gamma d\to\pi NN$ channel to the GDH integral
up to $E_\gamma=1.5$~GeV was
evaluated in the AFS approach~\cite{AFS04}, but the sensitivity
of obtained results to the $\gamma N\to\pi N$ operator
was not investigated in that work. An analysis of
Ref.~\cite{darwish07} shows that
the value of the integral at integration up to 350~MeV
manifests its considerable dependence on
the elementary operator. Because the energy domain from
threshold to 350~MeV provides only about a half of
the total contribution of this reaction to the deuteron GDH integral,
it is of high interest to recognize how sensitive the total integral is to
the $\gamma N\to\pi N$ operator. In the present paper, we evaluate
the integral up to a photon energy of $E_\gamma=1.65$~GeV and
investigate a question of possible theoretical uncertainties for its value.

The work is organized as follows. In Sec.~\ref{kinema},
the kinematic relations used in the calculations as well as definitions
for observables are  reviewed. A brief description of the
theoretical model and its ingredients is given in Sec.~\ref{theory}.
Section~\ref{results} contains results
for the unpolarized and helicity-dependent yields and cross sections and
their comparison  with the data available in the considered kinematic region.
Results of the evaluation of the $\gamma d\to\pi NN$ channel
contribution to the deuteron GDH integral
are also given in Sec.~\ref{results}.
In Sec.~\ref{conclusion}, we summarize  the main conclusions
and results.

\section{Kinematics}
\label{kinema}

Actual calculations of the reaction amplitude and observables are done
in the laboratory frame where
$k=(E_\gamma,\vec k),~p_d=(M,\vec 0),~p_\pi=(\varepsilon_\pi,\vec q),~p_1
=(\varepsilon_1,\vec p_1)$, and
$p_2=(\varepsilon_2,\vec p_2)$ are the four-momenta of the initial photon
and deuteron and the final pion and nucleons, respectively.
For the case when the final pion is detected, we
take as  independent kinematic  variables the
photon energy $E_\gamma$, the value of the pion momentum $q=|\vec q|$,
the pion polar angle $\Theta_\pi$, and the solid angle
$\Omega_{\vec  P_{NN}}$ ($\Theta_{\vec P_{NN}}$, $\phi_{\vec P_{NN}}$)
of the relative momentum $\vec P_{NN}$
of the final nucleon-nucleon pair.  These variables
totally determine the kinematics.

The unpolarized differential cross section is given by
\beqn
\label{dcs0}
\frac {d\sigma }{d\vec q d\Omega_{\vec  P_{NN}}}=f_{NN}
~~\frac 16
\sum _{m_2m_1\lambda m_d}
| \langle m_2m_1| T| \lambda m_d \rangle |^2,
\eeqn
where $\langle m_2m_1| T| \lambda m_d \rangle$ is the reaction amplitude,
and the phase space factor is
\beq
\label{fNN}
f_{NN}=\frac 1{(2\pi)^5}~\frac {m^2|{\vec P_{NN}}|}
      {4E_\gamma\varepsilon_\pi W_{NN}},
\eeq
with $W_{NN}=\sqrt{(p_1+p_2)^2}$ and $m$ is the nucleon mass.
Spin states of the two nucleons  and deuteron
are $m_2$, $m_1$,  and $m_d$, respectively, and they are chosen with respect
to the $z$ axis, which is defined by the photon momentum $\vec k$.
The symbol $\lambda$ stands for the photon helicity.
An extra factor of 1/2 must be included in the rhs of
Eq.~(\ref{fNN}) in the case of charged pion production.

We will be also interested in double polarized observables
for the polarized photon beam and polarized deuteron target.
These are the parallel, $P$, and antiparallel, $A$, cross sections
defined as
\beqn
\label{dcs_P0}
\frac {d\sigma_P }{d\vec q d\Omega_{\vec  P_{NN}}}&=&f_{NN}
~~\frac 12
\sum _{m_2m_1}
\Big(|\langle m_2m_1| T|+1+1\rangle |^2+|
\langle m_2m_1| T|-1-1\rangle |^2\Big),
\\
\label{dcs_A0}
\frac {d\sigma_A }{d\vec q d\Omega_{\vec  P_{NN}}}&=&f_{NN}
~~\frac 12
\sum _{m_2m_1}
\Big(|\langle m_2m_1| T|+1-1\rangle |^2+|
\langle m_2m_1| T|-1+1\rangle |^2\Big).
\eeqn

To obtain the semi-inclusive differential cross sections
$d\sigma/d\Omega_\pi$, $d\sigma_P/d\Omega_\pi$, and $d\sigma_A/d\Omega_\pi$,
the right-hand sides of Eqs.~(\ref{dcs0}), (\ref{dcs_P0}), and (\ref{dcs_A0})
have to be integrated
over $q$ and the solid angle $\Omega _{\vec P_{NN}}$, i.e.,
\beqn
\label{dcs}
\frac {d\sigma}{d\Omega_\pi}&=&\int_{q^{\rm min}}^{q^{\rm max}}\!\!\!\!f_{NN}~q^2dq
\int d\Omega_{\vec  P_{NN}}~\frac {d\sigma }{d\vec q d\Omega_{\vec  P_{NN}}},
\\
\label{dcs_P}
\frac {d\sigma_P}{d\Omega_\pi}&=&\int_{q^{\rm min}}^{q^{\rm max}}\!\!\!\!f_{NN}~q^2dq
\int d\Omega_{\vec  P_{NN}}~\frac {d\sigma_P }{d\vec q d\Omega_{\vec  P_{NN}}},
\\
\label{dcs_A}
\frac {d\sigma_A}{d\Omega_\pi}&=&\int_{q^{\rm min}}^{q^{\rm max}}\!\!\!\!f_{NN}~q^2dq
\int d\Omega_{\vec  P_{NN}}~\frac {d\sigma_A}{d\vec q d\Omega_{\vec  P_{NN}}}.
\eeqn
The upper and lower integration limits in the laboratory frame are
defined by the following relations
\beqn
\label{q_max}
q^{\rm max}=q^{\rm max}(\Theta_\pi)=&&\frac 1b
\left[ aE_\gamma z_\pi +(E_\gamma+M)\sqrt{a^2-b\mu^2}~\right] ,
\\
q^{\rm min}=q^{\rm min}(\Theta_\pi)=&{\rm max}\Big\{0,&\frac 1b
\left[ aE_\gamma z_\pi-(E_\gamma+M)\sqrt{a^2-b\mu^2}~\right]\Big\} ,
\label{q_min}
\eeqn
where $a=(W_{\gamma d}^2-4m^2+\mu^2)/2$,
$b=(E_\gamma+M)^2-E_\gamma^2z_\pi^2$, $z_\pi=\cos{\Theta_\pi}$,
and $\mu$ is the pion mass.

For the neutral channel, the emitted proton was registered in
the experiment~\cite{ahrens10}. In this case we chose as
independent kinematic variables the photon energy $E_\gamma$,
the value of the proton momentum $p_p=|\vec p_p|$,
the proton polar angle $\Theta_p$, and the solid angle
$\Omega_{\vec  P_{\pi n}}$ ($\Theta_{\vec P_{\pi n}}$, $\phi_{\vec P_{\pi n}}$)
of the relative momentum $\vec P_{\pi n}$ of the final meson-neutron pair.
Equations analogous to Eqs.~(\ref{q_max}) and (\ref{q_min}) are now of the form
\beqn
\label{p_max}
p^{\rm max}_p=p^{\rm max}_p(\Theta_p)=&&\frac 1B
\left[ AE_\gamma z_p+(E_\gamma+M)\sqrt{A^2-Bm^2}~\right] ,
\\
p^{\rm min}_p=p^{\rm min}_p(\Theta_p)=&{\rm max} \Big\{0,&\frac 1B
\left[ AE_\gamma z_p-(E_\gamma+M)\sqrt{A^2-Bm^2}~\right]\Big\},
\label{p_min}
\eeqn
where $A=(W_{\gamma d}^2-2m\mu - \mu^2)/2$,
      $B=(E_\gamma+M)^2-E_\gamma^2z_p^2$, and $z_p=\cos{\Theta_p}$.

Only a fraction of the total interval $[p^{\rm min}_p,p^{\rm max}_p]$
was accessed in conditions of the experiment~\cite{ahrens10}.
There were registered the protons with
the absolute value of the momentum $p_p\geq p^{\rm min}_{\rm exp}$, where
\beq
p^{\rm min}_{\rm exp}=p^{\rm min}_{\rm exp}(\Theta_p)=
[210+0.01\times(\Theta_p-90^\circ)^2]~{\rm MeV}/c,
\eeq
so that the differential $\pi^0np$ yields $Y$, as they were called
in Ref.~\cite{ahrens10}, were measured rather than
the differential cross sections.
In analogy to Eqs.~(\ref{dcs})--(\ref{dcs_A}) they are defined as
\beqn
\label{Y0}
\frac {dY}{d\Omega_p}&=&\int_{p^{\rm min}_{\rm exp}}^{p_p^{\rm max}}\!\!\!\!
f_{\pi n}~p_p^2~dp_p\int d\Omega_{\vec  P_{\pi n}}~\frac 16
\sum _{m_pm_n\lambda m_d}
| \langle m_pm_n| T| \lambda m_d \rangle |^2,
\\
\label{Y_P}
\frac {dY_P}{d\Omega_p}&=&\int_{p^{\rm min}_{\rm exp}}^{p_p^{\rm max}}\!\!\!\!
f_{\pi n}~p_p^2~dp_p\int d\Omega_{\vec  P_{\pi n}}~\frac 12
\sum _{m_pm_n}
\Big(|\langle m_pm_n| T|+1+1\rangle |^2+|
\langle m_pm_n| T|-1-1\rangle |^2\Big),
\\
\label{Y_A}
\frac {dY_A}{d\Omega_p}&=&\int_{p^{\rm min}_{\rm exp}}^{p_p^{\rm max}}\!\!\!\!
f_{\pi n}~p_p^2~dp_p\int d\Omega_{\vec  P_{\pi n}}~\frac 12
\sum _{m_pm_n}
\Big(|\langle m_pm_n| T|+1-1\rangle |^2+|
\langle m_pm_n| T|-1+1\rangle |^2\Big),
\eeqn
where
\beq
f_{\pi n}=\frac 1{(2\pi)^5}~\frac {m^2|{\vec P_{\pi n}}|}
      {4E_\gamma\varepsilon_p W_{\pi n}},
\eeq
and $W_{\pi n}=\sqrt{(p_\pi+p_n)^2}$.

\section{The formalism}
\label{theory}

\begin{figure*}[hbt]
\includegraphics[width=0.21\textwidth]{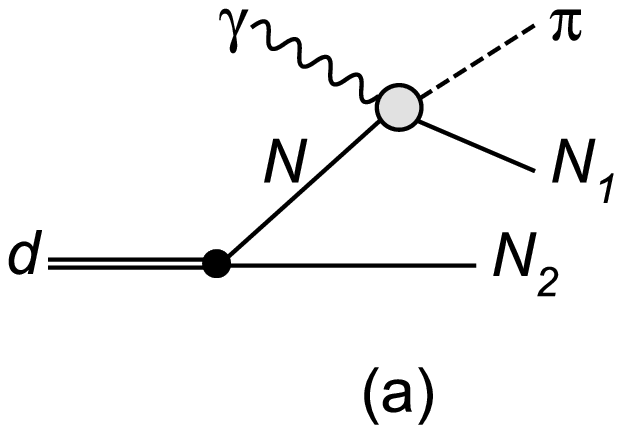}
\includegraphics[width=0.27\textwidth]{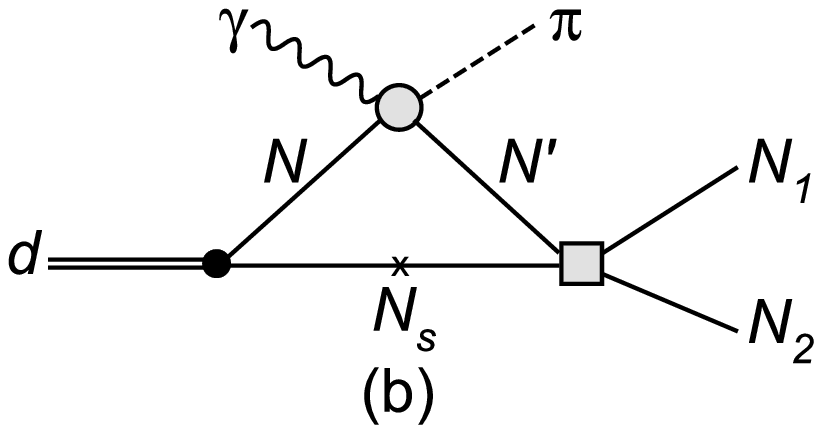}
\includegraphics[width=0.27\textwidth]{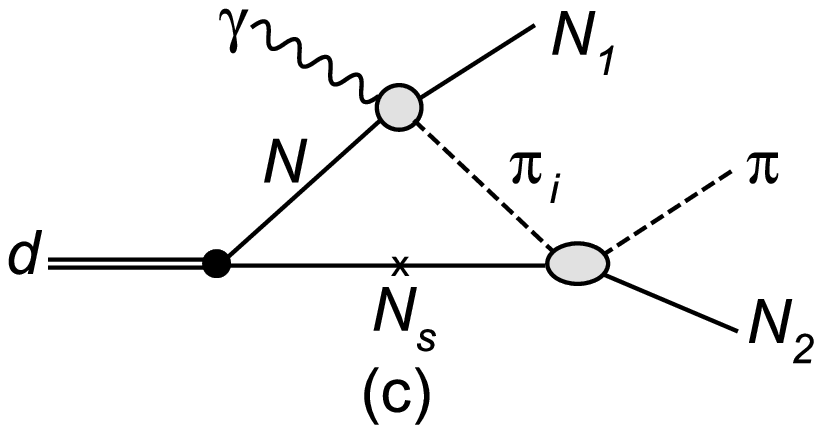}
\caption{
Diagrammatic representation of the $\gamma d\to \pi NN$ amplitude.
Diagram~(a) corresponds to IA.
Mechanisms with $NN$ and $\pi N$ rescattering in
the final state are shown in diagrams~(b) and (c), respectively.
Diagrams with the permutation $N_1\leftrightarrow N_2$ are not shown.
The spectator nucleon indicated by an ${\rm x}$ is on-shell.
 }
\label{diagrams}
\end{figure*}

In this section we briefly outline the formalism used
to evaluate the $\gamma d\to\pi NN$ reaction amplitude.
It was described in Refs.~\cite{lps96,lsw00,ls06}, to
which the reader is referred for more details.
The diagrammatic approach is exploited to calculate
the $\gamma d\to\pi NN$ reaction amplitude.
Because in this work we are not interested in the
threshold region, we reduce the set of diagrams in comparison
to that considered in Ref.~\cite{lsw00}.
Specifically, only contributions from the three diagrams
shown in Figs.~\ref{diagrams}(a)--(c) to the reaction amplitude
have been taken into account. Diagram~\ref{diagrams}(a)
is often refereed to as IA.
Sometimes it is called the pole diagram. Apart from it,
the diagrams~\ref{diagrams}(b) and (c) corresponding,
respectively, to $NN$ and $\pi N$ rescattering
in the final state have been considered in the present
calculation.

\begin{figure*}[hbt]
\includegraphics[width=0.27\textwidth]{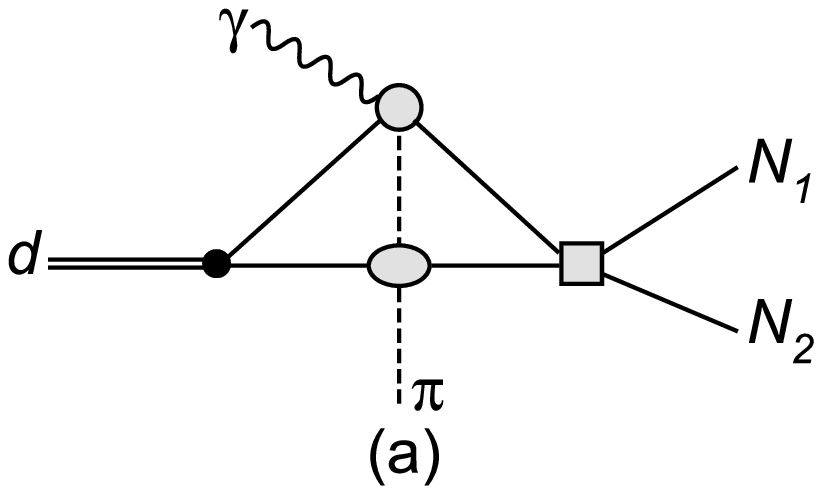}
\includegraphics[width=0.27\textwidth]{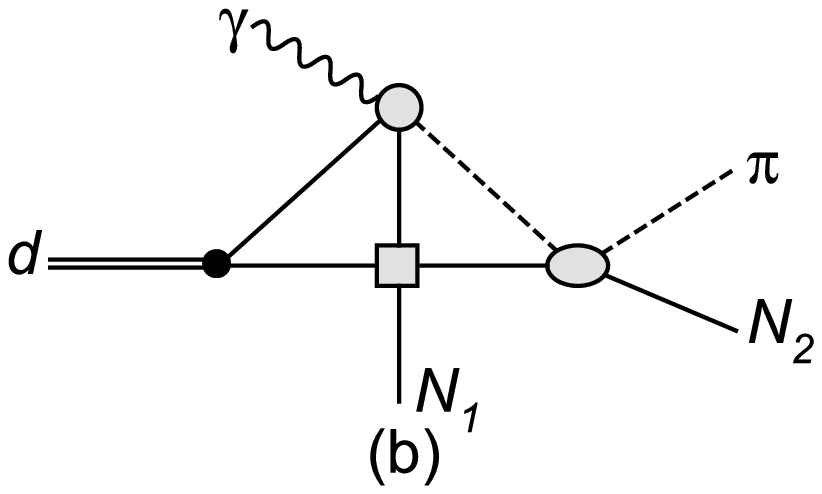}
\includegraphics[width=0.27\textwidth]{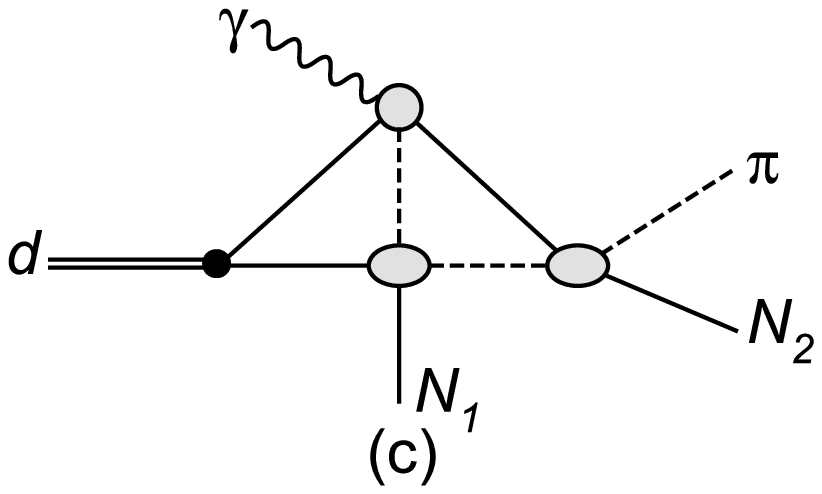}
\caption{
Two-loop diagrams of the reaction $\gamma d\to \pi NN$.
The permutation $N_1\leftrightarrow N_2$ is not shown.
 }
\label{2loops}
\end{figure*}

One remark should be made here.  A detailed analysis of
the helicity dependence of the $\vec \gamma\vec d\to\pi NN$
reaction up the $\Delta (1232)$ resonance
is presented in Ref.~\cite{darwish07}.
Special attention was given in that work to the near-threshold region.
Within the approach \cite{darwish07} the same set
of diagrams as presented in Fig.~\ref{diagrams} was taken into account.
One can expect such a model to be quite realistic
near threshold for the description of the charged channels.
But this is not obviously to be the case for the neutral channel.
As was shown in Ref.~\cite{lsw00}, a two-loop diagram as
in Fig.~\ref{2loops}(a)
which includes simultaneously  $\pi N$ rescattering
in the intermediate state and $np$ rescattering in the final state
has to be taken into account at threshold energies.
A similar mechanism is known to be
very important near threshold for coherent $\pi^0$
photoproduction on the deuteron
$\gamma d\to\pi^0d$~\cite{koch,bosted,benmer}.

The importance of diagram (a) in Fig.~\ref{2loops}
at threshold energies is that it contains,
together with the $np$ scattering amplitude,
a block with charged pion photoproduction from the nucleon.
The dominating threshold electric dipole amplitude $E_{0+}$
for the charged
channels is about 20 times larger in absolute numbers than that
for the neutral channels.
For analogous reasons we also expect that
a two-loop diagram~\ref{2loops}(b) with $NN$ rescattering
in the intermediate state and $\pi N$ rescattering in the final state
may also be important near threshold. Therefore, the fact that
the model including only three contributions shown
in Figs.~\ref{diagrams}(a)--(c) can be used
for the description of the $\gamma d\to\pi^0np$ reaction at threshold energies
is not obvious, and further study is needed to clarify this point.
There is one more two-loop diagram, shown in Fig.~\ref{2loops}(c),
which includes two blocks with $\pi N$ scattering. Because
the $\pi N$ scattering lengths are about two orders smaller
than those for $NN$ scattering, the role of this diagram
in the threshold region is expected to be much smaller than that of
diagrams~\ref{2loops}(a) and \ref{2loops}(b).

One can anticipate the effect from the above two-loop diagrams to decrease
together with increasing $E_\gamma$ when the mentioned enhancement
due to the charge pion exchange disappears. Nevertheless,
to be fully confident in the smallness of such contributions,
their explicit evaluations are needed. This task, however,
is expected to be very
difficult, because as we have learned evaluating diagram~\ref{2loops}(a)
in Ref.~\cite{lsw00},  practical calculations are extremely
time consuming especially when the inclusive processes are considered,
and, therefore, integrations over momenta of final particles have to
be carried out.
Note that in the case of the exclusive channels,
the kinematic regions exist where the
diagrams in Figs.~\ref{diagrams}(a)--\ref{diagrams}(c)
can be suppressed, and
other mechanisms, in particular, the two-loop
diagrams~\ref{2loops}(a)--\ref{2loops}(c) become important and
have to be taken into account (see Refs.~\cite{laget78,laget81}
for detailed discussions of this point).

Let us now write explicit expressions for the amplitudes corresponding
to diagrams~\ref{diagrams}(a)--\ref{diagrams}(c).
One has for the IA amplitude
\beq
\label{IA}
\langle m_2m_1
|T^{\rm IA}(\vec p_2,\vec p_1,\vec q;\vec k)|
\lambda m_d
\rangle =\sum _m
\Psi ^{m_d}_{m_2m}\left(\vec p_2 \right)
\langle m_1|
T_{\gamma N\to \pi N_1}(\vec p_1,\vec q;\vec k)
| \lambda m \rangle,
\eeq
where $\Psi ^{m_d}_{m_2m}$ is the deuteron wave function having
the form
\beqn
\label{Psi}
\Psi ^{m_d}_{m_2m_1}(\vec p)=\sum_{L=0,2}i^L
C^{1m_S}_{\frac 12 m_2\frac 12 m_1}C^{1m_d}_{1m_SLm_L}Y_L^{m_L}(\hat {\vec p})
u_L(p).
\eeqn
In Eq.~(\ref{Psi}), $Y^{m_L}_{L}({\hat {\vec p}})$
are the spherical harmonics, and
$C^{JM}_{J_1M_1J_2M_2}$ are the Clebsch-Gordan coefficients. The $S$
and $D$ wave function amplitudes $u_0(p)$ and $u_2(p)$ are taken
for the CD-Bonn potential~\cite{mach01}. One should emphasize
that calculations with another realistic version of the nucleon-nucleon
potential, namely, with the Nijmegen model~\cite{Nijmegen},
have led to essentially the same results, so we have found no sensitivity
of our predictions to the choice of a potential. Note that in Eq.~(\ref{IA})
and in the amplitudes that follow, we do not write explicitly
those corresponding to the permutation
$N_1\leftrightarrow N_2$, but they are included in the calculations.

The symbol $T_{\gamma N\to \pi N} $ in Eq.~(\ref{IA}) stands for
the elementary operator of pion photoproduction on the nucleon.
As is explained in detail in Ref.~\cite{ls06} (see, also Ref.~\cite{ahrens10}),
ambiguities in predictions for observables
in the $\gamma d\to\pi NN$ reaction stem mainly from the manner
in which $T_{\gamma N\to \pi N}$ is embedded into the deuteron.
In all recent calculations~\cite{FixAr05,ls06,darwish07}, this
operator is taken in the on-shell form and it is parameterized using
either the multipole analyses SAID~\cite{SAID} and MAID~\cite{MAID} or
the effective Lagrangian approach (ELA)~\cite{ELA}. Although the models
are close in their predictions for observables
in the reaction $\gamma N\to \pi N$,
they are not identical. It is evident that this difference
will manifest itself in results of the evaluation of Eq.~(\ref{IA}).
To estimate ambiguities caused by different choices of
the on-shell operator  $T_{\gamma N\to \pi N}$,
we perform our calculations with the latter taken from the recent
multipole analyses
SAID~\cite{SAID} (solution SP09K) and MAID07~\cite{MAID07}.

Furthermore, the on-shell operator $T_{\gamma N\to \pi N} $ depends
on four invariant amplitudes~\cite{CGLN} for which there exist different
options that are equivalent in the on-shell case. This equivalence is
broken when one or both nucleons are off their mass shells.
Just such a situation takes place in Eq.~(\ref{IA}) where the nucleon $N$
is off its mass shell. In this work, we evaluate
the operator $T_{\gamma N\to \pi N}$ with two sets of the
invariant amplitudes, $A_i$ and $A_i'$, as they are called
in Ref.~\cite{ls06} and where their definition can be found.
At least, this allows one to estimate the
possible uncertainties in the predicted observables due to off-shell
effects in $T_{\gamma N\to \pi N}$.

The matrix element corresponding to  Fig.~\ref{diagrams}(b) reads
\beqn
\langle m_2m_1| T^{NN}(\vec p_2,\vec p_1,\vec q;\vec k)
| \lambda m_d\rangle &=&
m\int \frac {d^3{\vec p}_s}{(2\pi)^3}
\frac 1{p^2_{\rm out}-p^2_{\rm in}+i0}
\nonumber
\\
&&
\times\sum_{m_sm'}\langle{\vec p}_{\rm out},m_2m_1| T_{NN}(E)|{\vec p}_{\rm in},
 m_sm'\rangle
\langle m_sm'| T^{\rm IA}(\vec p_s,\vec p\,',\vec q;\vec k)|\lambda m_d
\rangle.
\label{NNresc}
\eeqn
Here $T^{\rm IA}$ is the IA amplitude given by  Eq.\ (\ref{IA}).
The relative momenta  of the $N_2N_1$ pair after and before scattering
are $\vec p_{\rm out}=(\vec p_2-\vec p_1)/2$ and $\vec p_{\rm in}={\vec
p}_s-(\vec p_2+\vec p_1)/2$, respectively, and $E=p_{\rm out}^2/m$.
The half-off-shell $NN$ scattering matrix $T_{NN}$ has been obtained by
solving an equation of the Lippmann-Schwinger type for the
CD-Bonn potential~\cite{mach01}. All states with
the total angular momentum $J\leq 2$ have been retained in $T_{NN}$.

In the evaluation of Eq.~(\ref{NNresc}) there is another ambiguity
due to different prescriptions for the determination
of the total invariant energy entering the $\gamma N$ vertex.
As was proposed in Refs.~\cite{laget78,laget81},
we define the latter as if the spectator nucleon $N_s$
was on its mass-shell.
Other prescriptions can also be found in
the literature. For instance, in Ref.~\cite{FixAr05}
the active nucleon $N$ is put to be on its mass-shell.
We have evaluated Eq.~(\ref{NNresc}) with these two
prescriptions and found only at most a 2\% variation for
all observables considered in the next section. This means that
the above ambiguity can be safely disregarded.

The matrix element corresponding to Fig.~\ref{diagrams}(c)
with $\pi N$ rescattering in the final state is
\beqn
\langle m_2m_1| T^{\pi N}(\vec p_2,\vec p_1,\vec q;\vec k)
| \lambda m_d\rangle &=&
\int \frac {d^3{\vec p}_s^{\,*}}{(2\pi)^3}~
\frac {\varepsilon_2^*+\varepsilon_s^*}{2W_{\pi N_2}}~
\frac 1{{\vec p_2}^{\,*2}-{\vec p_s}^{\,*2}+i0}~F^2_{\pi N}(|\vec q_{\pi_i}|)
\nn
&&
\times\sum_{m_s}\Big[
\langle{\vec p}\,^*_2,m_2| T^0_{\pi N}(W_{\pi N_2})|{\vec p}_s^{\,*},
 m_s\rangle
\langle m_sm_1| T^{\rm IA}_{\pi^0 N_1}(\vec p_s,\vec p_1,\vec q_{\pi_i};\vec k)|
\lambda m_d\rangle
\label{piNresc}
\\
&&
~~~~~-\langle{
\vec p}\,^*_2,m_2| T^{\rm ch}_{\pi N}(W_{\pi N_2})|{\vec p}_s^{\,*},
 m_s\rangle
\langle
m_sm_1| T^{\rm IA}_{\pi^{\rm ch} N_1}(\vec p_s,\vec p_1,\vec q_{\pi_i};\vec k)|
\lambda m_d\rangle \Big],
\nonumber
\eeqn
where the asterisk denotes variables in the $\pi N_2$ center-of-mass (c.m.)
frame, i.e.
${\vec p}_s^{\,*}=-{\vec q}_\pi^{\,*}$, where
${\vec p}_s^{\,*}$ and ${\vec q}_\pi^{\,*}$
are the c.m. initial nucleon and pion
momenta, respectively, in the $\pi N$ scattering vertex.
Analogously, ${\vec p}_2^{\,*}=-{\vec q}^{\,*}$,
where  ${\vec p}_2^{\,*}$ and ${\vec q}^{\,*}$ are the c.m. final nucleon and pion
momenta, respectively. The momenta with the asterisk and without it are
related to each other through a boost transformation
with the velocity $(\vec q + \vec p_2)/(\varepsilon_\pi + \varepsilon_2)$.
We take the spectator
nucleon $N_s$ to be on its mass shell, i.e.
$\varepsilon_s^*=\sqrt{p^{*2}_s+m^2}$.  The total energy $W_{\pi N_2}$
of the $\pi N_2$ pairs is
$W_{\pi N_2}=\sqrt{(q+p_2)^2}=\varepsilon_2^*+\varepsilon_{\pi}^*=
\sqrt{{\vec p_2}^{\,*2}+m^2}+\sqrt{{\vec q}^{\,*2}+\mu^2}$.

The half-off-shell $\pi N$ scattering matrix $T_{\pi N}$
has been obtained by solving the Lippmann-Schwinger
equation for a separable energy-dependent $\pi N$ potential
built in Ref.~\cite{nozawa}. We have checked that the results remain
essentially the same as the $T_{\pi N}$ found
from a meson-exchange model \cite{hung94,hung01}.
The factor $F^2_{\pi N}(|\vec q_{\pi_i}|)$ in Eq.~(\ref{piNresc})
is introduced
to take into account the off-shell nature of the intermediate meson $\pi_i$.
In accordance with Ref.~\cite{nozawa}, we choose the form factor $F_{\pi N}(q)$
of the $\pi NN$ vertex to be of the monopole form
\beq
F_{\pi N}(q)=\frac { \Lambda^2_\pi}{\Lambda^2_\pi +q^2},
\eeq
with a cutoff parameter $\Lambda_\pi=650$ MeV/{\it c}.

The superscript indices ``$0$" and ``$\rm ch$" in the $T_{\pi N}$ amplitude
in Eq.~(\ref{piNresc}) stand for
the neutral, $\pi^0N\to\pi^0N_1$, or charge exchange,  $\pi^{\rm ch}N\to\pi^0N_1$,
channels, respectively.
The IA amplitudes for $\pi^0$ photoproduction, $T^{\rm IA}_{\pi^0 N_1}$, and
for charged pion photoproduction, $T^{\rm IA}_{\pi^{\rm ch} N_1}$,
are given by Eq.~(\ref{IA}).

\section{Results and discussion}
\label{results}

\subsection{Unpolarized yield and cross section}
\label{sub:unpolar}

We begin the discission of the results with the unpolarized yield
and cross section. Special emphasize is given to their comparison
with the very recent experimental data from
Refs.~\cite{ahrens10,ahrens09,ahrens06}.
A detailed comparison with older data can be found in Ref.~\cite{ls06}.
As a rule, we will not compare our predictions with the results of
other theoretical approaches~\cite{FixAr05,darwish07,schwamb10}
because this has already been done in Ref.~\cite{ahrens10}.

\begin{figure*}[hbt]
\includegraphics[width=0.6\textwidth]{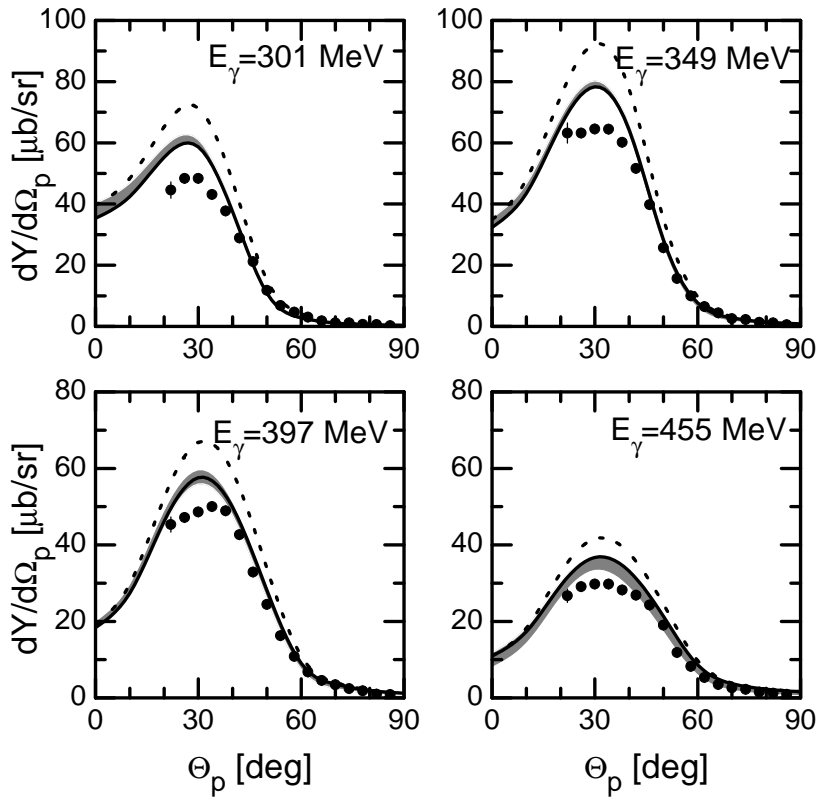}
\caption{
Unpolarized differential yield
for the $\gamma d\to\pi^0np$ reaction at four selected energies.
Dotted and solid curves are obtained
in IA and the full model, respectively,
with the MAID07 analysis and the amplitudes $A_i$.
Shaded areas are our predictions
with different parameterizations of the elementary $\gamma N\to\pi N$
amplitude  as described in the text. Data are from
Ref.~\cite{ahrens10}. Only statistical errors are shown.
 }
\label{pi0UNPOL}
\end{figure*}

Results for the unpolarized differential yield (\ref{Y0})
for the $\gamma d\to\pi^0np$ reaction at four selected energies
from 301 to 455 MeV with a step of about 50 MeV
are shown in Fig.~\ref{pi0UNPOL}.
One can see a sizable effect from FSI, which is mainly
due to $np$ FSI.
The yield is reproduced very well at the proton angles
$\Theta_p\geq 40^\circ$.
However, despite of the strong reduction of the unpolarized
differential yield due to FSI, the full model clearly overestimates
the data in the peak position.
An analogous feature was found also in Ref.~\cite{ls06}
where after integration over complete acceptance, i.e. with
$p^{\rm min}_p$ given by Eq.~(\ref{p_min}), we observed the predictions
to lie well above data from Refs.~\cite{krusche99,siodl01}.
Similar overestimation, although in some variation of size,
is present in all recent calculations~\cite{darwish03,FixAr05,schwamb10}.

It is difficult to find a reasonable explanation for this disagreement
in the framework of the present model.
First, reasons for a possible failure
of the IA calculation are not apparent.
The contribution of IA to the total cross section stems from
the quasi-free $\pi^0 N$ process, which is evaluated very reliably.
Second, as explained in Ref.~\cite{ls06}, the $NN$ FSI contribution
comes mainly from the kinematic domains where nucleons $N$ and $N'$
in Fig.~\ref{diagrams}(b) are close to their mass shells so that
the on-shell parameterization of the elementary operator
$T_{\gamma N\to \pi N}$ can be used in Eq.~(\ref{NNresc}).
Because modern  parameterizations  of $T_{\gamma N\to \pi N}$ and $T_{NN}$
are firmly established, we think that the effect of $NN$ FSI is
well under control. Therefore, the noticeable disagreement with the data
cannot be explained by an underestimation of the $NN$ FSI contribution.
We also suppose that $\pi N$ FSI is evaluated quite reasonably
and it can not be responsible for the disagreement, especially if
one takes into account that the $\pi N$ FSI effect is very small.

As discussed in Refs.~\cite{ahrens10,schwamb10}, a possible way
to resolve the problem is to take into account the absorption
of the produced pions by the pair of nucleons.
Rough estimations of the absorption effect performed
in Ref.~\cite{ahrens10} show that it can comprise more than 10\%
of the total cross section, thus leading to visible attenuation
of the production rate.

Model uncertainties of the predictions stemming from the
choice of the elementary operator $T_{\gamma N\to \pi N}$
are shown in the shaded areas in Fig.~\ref{pi0UNPOL}.
These areas have been generated, first, by results with the MAID07
and SAID (solution SP09K) multipole analyses and, second,
with two sets of the invariant amplitudes, $A_i$ and $A_i'$,
as described in the previous section. One can see that
the model dependence is quite small, and even when taken into account,
it does not resolve the problem with the description
of the data at $\Theta_p\leq40^\circ$.

\begin{figure*}[hbt]
\includegraphics[width=0.6\textwidth]{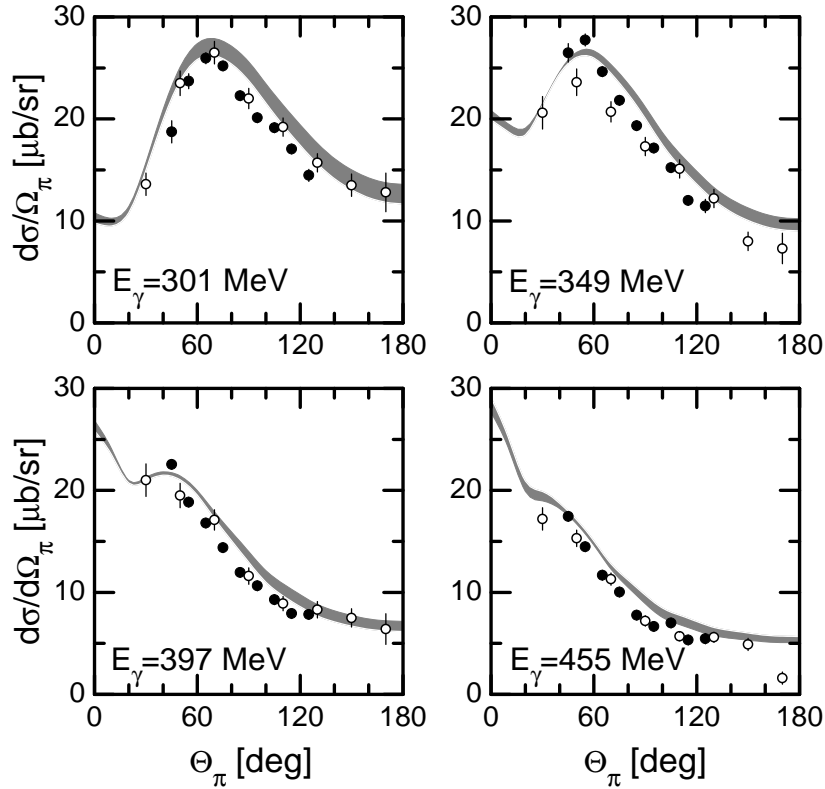}
\caption{
Unpolarized differential cross section
for the $\gamma d\to\pi^-pp$ reaction at four selected energies.
Shaded areas are our predictions
with different parameterizations of the elementary $\gamma N\to\pi N$
amplitude. Data are from
Refs.~\cite{ahrens10} ($\bullet$) and \cite{benz73} ($\circ$).
Only statistical errors are shown.
 }
\label{pimnUNPOL}
\end{figure*}

Predictions for the unpolarized differential cross section
in the $\gamma d\to\pi^-pp$ reaction are compared in Fig.~\ref{pimnUNPOL}
to the data from Refs.~\cite{ahrens10,benz73} at the same
energies as in Fig.~\ref{pi0UNPOL}. The agreement is quite
reasonable, although the slight overestimation of the data
is seen. The model dependence of the predictions is
mainly due to the choice of the multipole analysis and
is noticeable only at the lowest energy. It is compatible with
the data errors of Ref.~\cite{benz73}. Together with
increasing $E_\gamma$ this dependence diminishes and
becomes quite small above 350 MeV.

\begin{figure*}[hbt]
\includegraphics[width=0.6\textwidth]{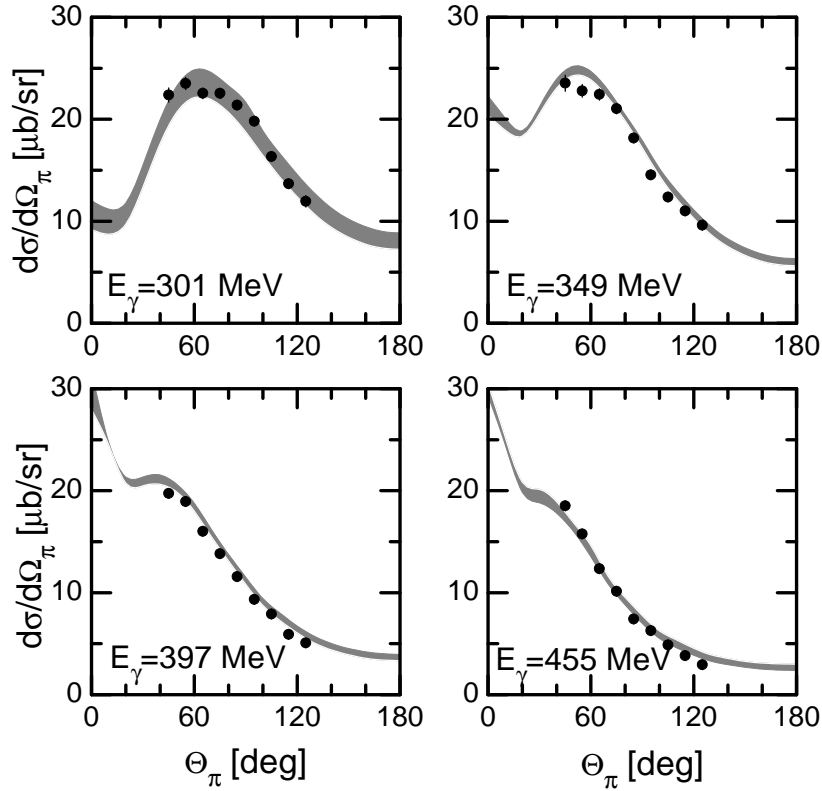}
\caption{
Same as Fig.~\ref{pimnUNPOL}, but for the $\gamma d\to\pi^+nn$ reaction.
 }
\label{piplUNPOL}
\end{figure*}

As is seen in Fig.~\ref{piplUNPOL}, the agreement
of the predictions and the unpolarized $\pi^+nn$ data
is almost perfect. Again, the visible dependence of the results
on the choice of the multipole analysis occurs only at 301 MeV.

\begin{figure*}[hbt]
\includegraphics[width=0.33\textwidth]{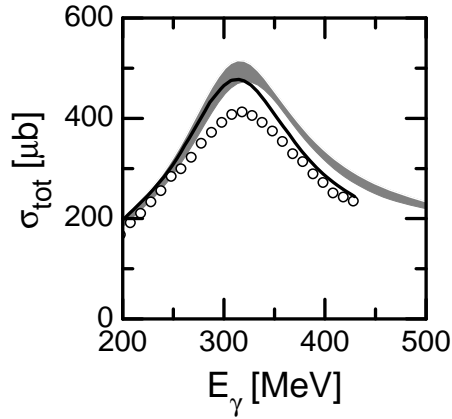}
\caption { Unpolarized total cross section for the charged
channels $\gamma d \to\pi^{\pm} NN$. Shaded area is our
predictions with different parameterizations of the elementary
$\gamma N\to\pi N$ amplitude. Solid curve is the AFS
result~\cite{AFS04}. Data are from Ref.~\cite{ahrens06}. }
\label{stot_charged}
\end{figure*}

Taking into account the results for the unpolarized
differential cross sections in the charged channels,
one could expect the sum of the total cross sections
for these two channels to be in reasonable agreement with
the data but with a possible slight overestimation because
of the above  overestimation in the $\pi^-pp$ reaction.
However, as is seen in Fig.~\ref{stot_charged}, the deviation
from the data of the GDH and A2 collaborations~\cite{ahrens06}
at $E_\gamma \geq$ 300 MeV is more serious.
At present, we are not aware of reasons for this disagreement.
Figure~\ref{stot_charged} shows also that our results are
in good agreement with these of AFS up
to the peak position at about 320 MeV,   but at higher energies
our cross sections overestimate the AFS results.

\subsection{Polarized yield and cross section}
\label{sub:polar}

\begin{figure*}[hbt]
\includegraphics[width=0.7\textwidth]{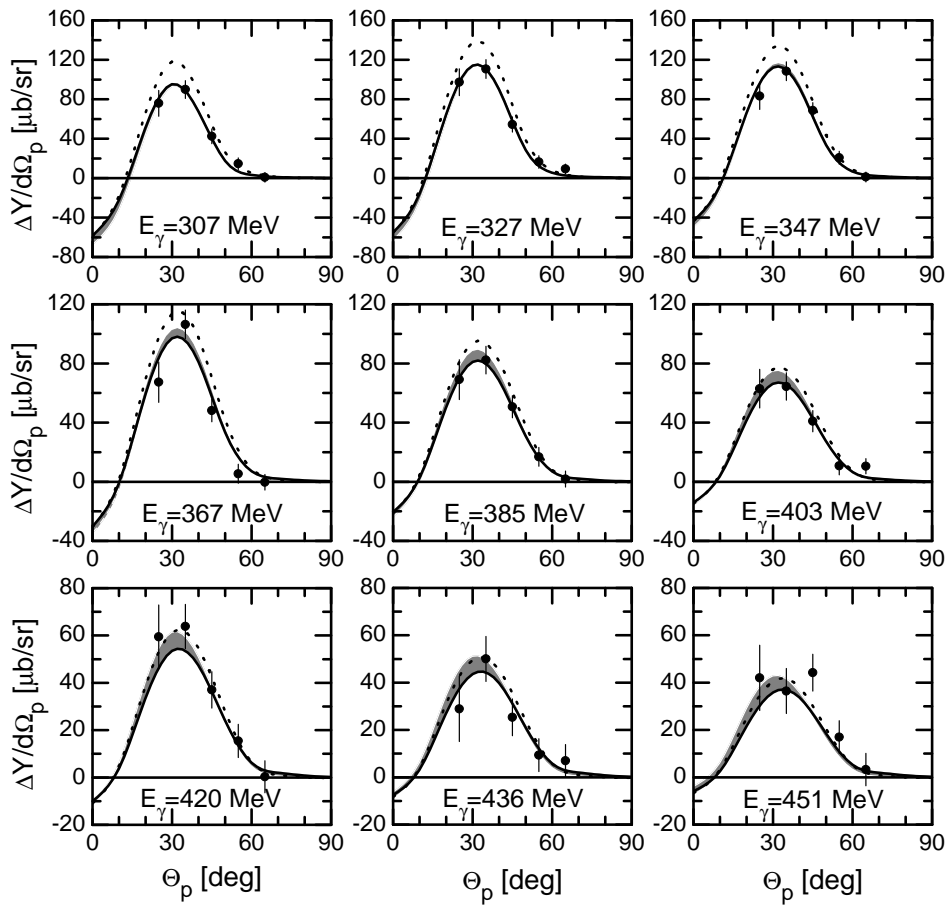}
\caption{
Helicity-dependent differential yield difference
$\Delta Y /d\Omega_p$ for the $\gamma d\to\pi^0np$ reaction.
Dotted and solid curves are obtained
in IA and the full model, respectively,
with the amplitudes $A_i$ and the MAID07 analysis.
Shaded areas are our predictions
with different parameterizations of the elementary $\gamma N\to\pi N$
amplitude. Data are from Ref.~\cite{ahrens10}.
Only statistical errors are shown.
 }
\label{pi0POL}
\end{figure*}

Next we discuss the polarized yield and cross section.
Figure~\ref{pi0POL} shows the helicity-dependent yield difference
\beq
\label{dY}
\frac {\Delta Y  }{d\Omega_p}=
\frac {      dY_P}{d\Omega_p}-
\frac {      dY_A}{d\Omega_p}
\eeq
for the $\pi^0np$ channel. Similar to the unpolarized yield, one
can see a sizable reduction of $\Delta Y/d\Omega_p$ in IA due
to FSI. However,  the striking  distinction from the unpolarized case
is that now the model reasonably reproduces all the data
not only those at $\Theta_p\geq 40^\circ$.
This might indicate that the above overestimation
in the unpolarized yield is due to an overestimation
in both $dY_P/d\Omega_p$ and $dY_A/d\Omega_p$,
but it disappears in the difference $\Delta Y/d\Omega_p$.

It is seen in Fig.~\ref{pi0POL} that the dependence of $\Delta Y /d\Omega_p$
on the elementary operator is weak below 350 MeV, but it becomes
quite visible at higher energies.

\begin{figure*}[hbt]
\includegraphics[width=0.7\textwidth]{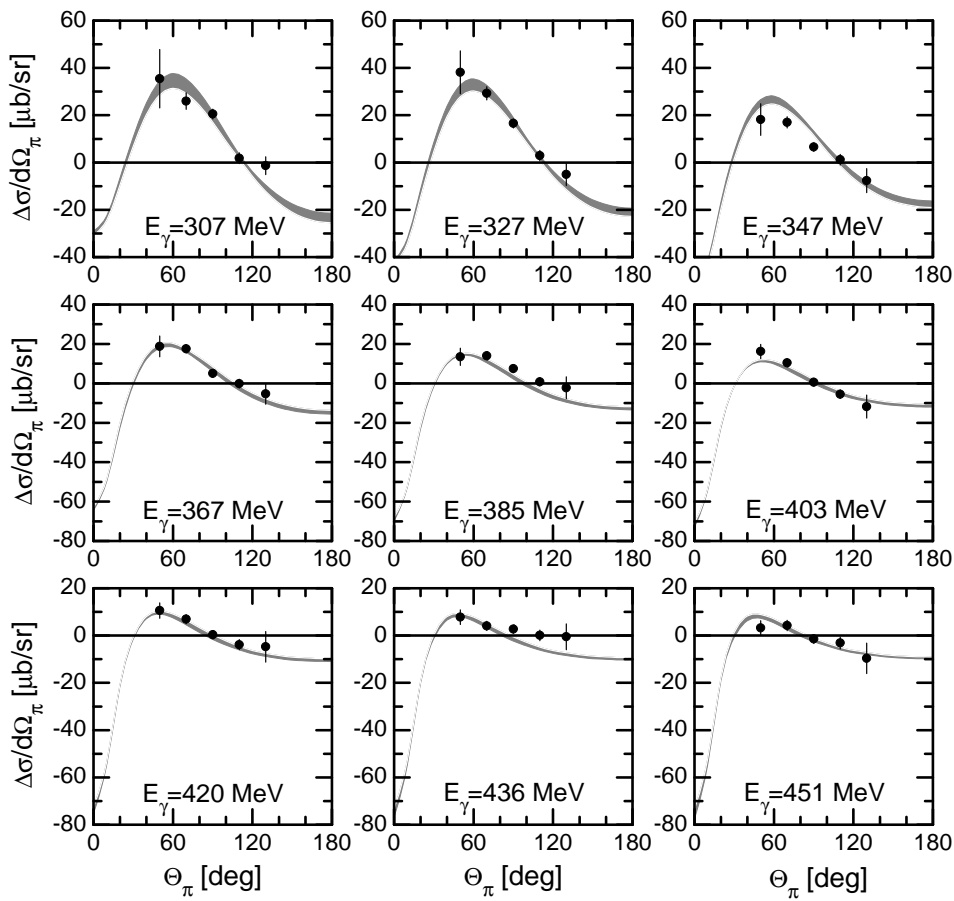}
\caption{
Helicity-dependent differential cross section difference
$\Delta \sigma /d\Omega_\pi$
for the $\gamma d\to\pi^-pp$ reaction. Shaded areas are our predictions
with different parametrizations of the elementary $\gamma N\to\pi N$
amplitude. Data are from Ref.~\cite{ahrens10}.
Only statistical errors are shown.
 }
\label{pimnPOL}
\end{figure*}

\begin{figure*}[hbt]
\includegraphics[width=0.7\textwidth]{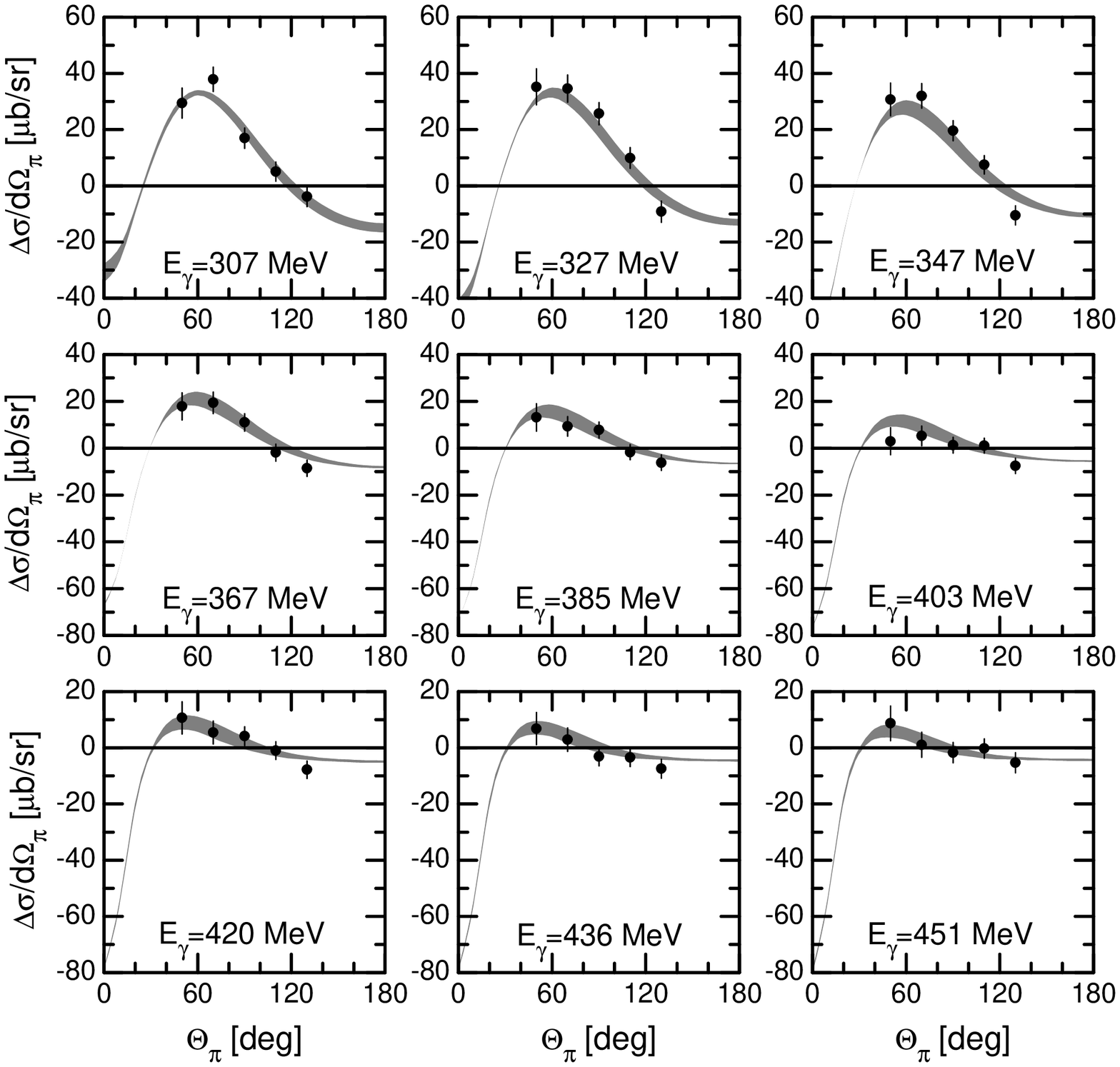}
\caption{
Same as Fig.~\ref{pimnPOL}, but for the $\gamma d\to\pi^+nn$ reaction.
 }
\label{piplPOL}
\end{figure*}

Figures~\ref{pimnPOL} and~\ref{piplPOL} show the helicity-dependent
differential cross section difference
\beq
\label{dsigma}
\frac {\Delta \sigma  }{d\Omega_\pi}=
\frac {      d\sigma_P}{d\Omega_\pi}-
\frac {      d\sigma_A}{d\Omega_\pi}
\eeq
for the $\pi^-pp$ and $\pi^+nn$ channels, respectively.
There is a good agreement with our results within
the uncertainties of the theoretical predictions.
Only a few $\pi^-pp$ data points are underestimated
at $\Theta_\pi=125^\circ$. The above uncertainties
are quite significant only in the $\pi^+nn$ channel
and just in the angular region overlapped in the
experiment~\cite{ahrens10}.

\begin{figure*}[hbt]
\includegraphics[width=0.33\textwidth]{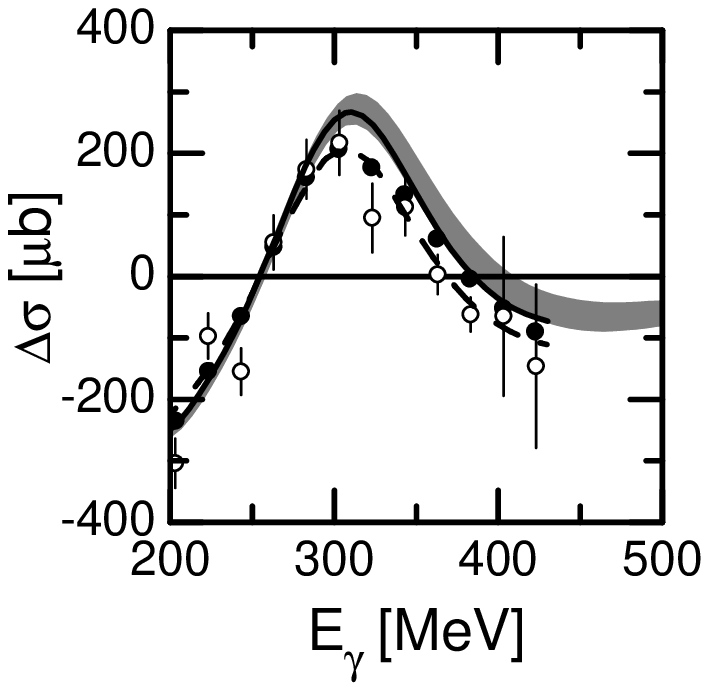}
\caption
{
Difference $\Delta\sigma$ for
the charged channels $\gamma d \to\pi^{\pm} NN$.
Shaded area is our predictions
with different parametrizations of the elementary $\gamma N\to\pi N$
amplitude.
Solid and dashed curves are the AFS~\cite{AFS04}
and Schwamb~\cite{schwamb10} results, respectively.
Data are from
Refs.~\cite{ahrens06} ($\circ$) and \cite{ahrens09} ($\bullet$).
Only statistical errors are shown.
}
\label{stotPmA}
\end{figure*}

An integration of Eq.~(\ref{dsigma}) over the solid angle $\Theta_\pi$
gives the difference $\Delta\sigma=\sigma_P-\sigma_A$ of
the total cross sections.
This difference for the semi-exclusive channels $\gamma d\to\pi^\pm NN$
and $\gamma d\to\pi^0X$ has been measured
in Refs.~\cite{ahrens06,ahrens09} at energies from 200 to 430 MeV.
Because we do not have a model for the coherent reaction $\gamma d\to\pi^0d$,
we are able to make predictions for the second channel.
Our results for the $\pi^\pm NN$ channel are shown in Fig.~\ref{stotPmA}.
Within theoretical uncertainties, the model satisfactory describes the data
of Ref.~\cite{ahrens06} below 300 MeV, but there exists noticeable
overestimation at higher energies, i.e. one observes the situation analogous
to that for the unpolarized cross section.
Reasonable agreement with the data of Ref.~\cite{ahrens09}
is achieved except for two points in the peak region.
Our results are in quite good agreement with the AFS predictions
but overestimate these from Ref.~\cite{schwamb10} at $E_\gamma \agt 280$ MeV.

Having measured the unpolarized yield and cross section as well as
$\Delta Y  /d\Omega_p$ and $\Delta \sigma / d\Omega_\pi$,
the authors of Ref.~\cite{ahrens10} have attempted to separate the
parallel and antiparallel components of the yield and cross section.
To do this, they used the follows relations:
\beqn
\label{ds}
      \frac { d\sigma  }{d\Omega}&=&\frac 13
\Big(2\frac {d\sigma_\parallel}{d\Omega}+\frac {d\sigma_0}{d\Omega}\Big),
\\
\label{T20}
T_{20}\frac { d\sigma  }{d\Omega}&=&\frac {\sqrt{2}}3
\Big(\frac {d\sigma_\parallel}{d\Omega}-\frac {d\sigma_0}{d\Omega}\Big),
\eeqn
where $\sigma_\parallel$ and $\sigma_0$ are the components of the cross
section corresponding to the deuteron spin states
$m_d=\pm 1$ and $m_d=0$, respectively, and $T_{20}$ is
the tensor target asymmetry~\cite{ArFix05,ls06}.
Note that the component $\sigma_\parallel$
is the average of the parallel and antiparallel cross sections
\beq
\label{ds_paral}
      \frac { d\sigma_\parallel}{d\Omega}=\frac 12
\Big(\frac {d\sigma_P}{d\Omega}+\frac {d\sigma_A}{d\Omega}\Big).
\eeq
It is follows
from Eqs.~(\ref{ds}) and (\ref{T20}) that
\beq
\label{ds_T20}
 \frac { d\sigma  }{d\Omega}=\frac {\sqrt{2}}{\sqrt{2}+T_{20}}~
 \frac { d\sigma_\parallel}{d\Omega}.
\eeq

Evaluations of $T_{20}$ in Refs.~\cite{FixAr05,ls06} showed that
in the angular regions accessed in conditions
of the experimental setup of Ref.~\cite{ahrens10}, i.e.
at $\Theta_\pi >20^\circ$, the absolute value of $T_{20}$
does not exceed  $0.05$. Therefore, one has the following approximate
relations  for $\Theta_\pi >20^\circ$:
\beq
\label{dsparal}
\frac { d\sigma            }{d\Omega} \approx
\frac { d\sigma_\parallel  }{d\Omega},\quad
\frac { d\sigma_\parallel  }{d\Omega} \approx
\frac { d\sigma_0          }{d\Omega}.
\eeq
Using Eqs.~(\ref{dsigma}), (\ref{ds}), (\ref{ds_paral}),
and (\ref{dsparal}) one obtains
\beq
\label{dsP/A}
 \frac { d\sigma_{P/A}  }{d\Omega} \approx
 \frac {d\sigma}{d\Omega}\pm  \frac 12 \frac {\Delta\sigma}{d\Omega}.
\eeq
It should be noted that the semi-inclusive asymmetry $T_{20}$ resulting from
the integration of the corresponding right-hand sides
over the kinematic regions appearing
in Eqs.~(\ref{Y0})--(\ref{Y_A}) was not evaluated
in Refs.~\cite{FixAr05,ls06}.
However, we have checked that such an integration gives $|T_{20}|\leq 0.08$
at $25^\circ\leq\Theta_p\leq 65^\circ$ and, therefore, the relations similar
to Eqs.~(\ref{dsparal}) and (\ref{dsP/A}) are valid
for the measured yields as well.

\begin{figure*}[hbt]
\includegraphics[width=0.7\textwidth]{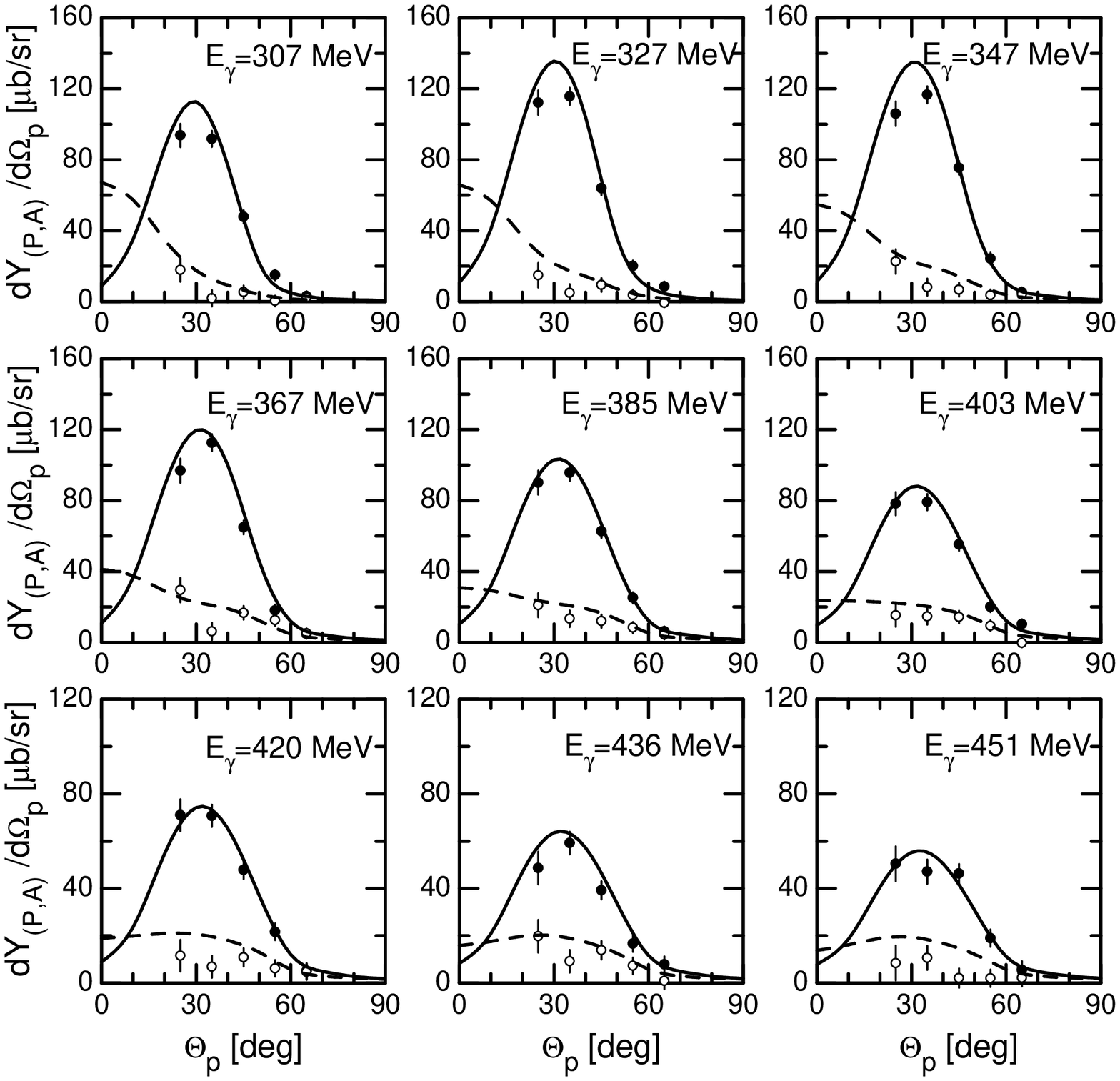}
\caption{
Helicity-dependent differential yields
$dY_P /d\Omega_p$ and $dY_A /d\Omega_p$
for the $\gamma d\to\pi^0np$ reaction. Solid and dashed curves are
our predictions for $dY_P /d\Omega_p$ and $dY_A /d\Omega_p$,
respectively, with the amplitudes $A_i$ and the MAID07 analysis.
Data are from Ref.~\cite{ahrens10} for the parallel ($\bullet$)
and antiparallel ($\circ$) components.
Only statistical errors are shown.
 }
\label{dY_PA}
\end{figure*}

\begin{figure*}[hbt]
\includegraphics[width=0.7\textwidth]{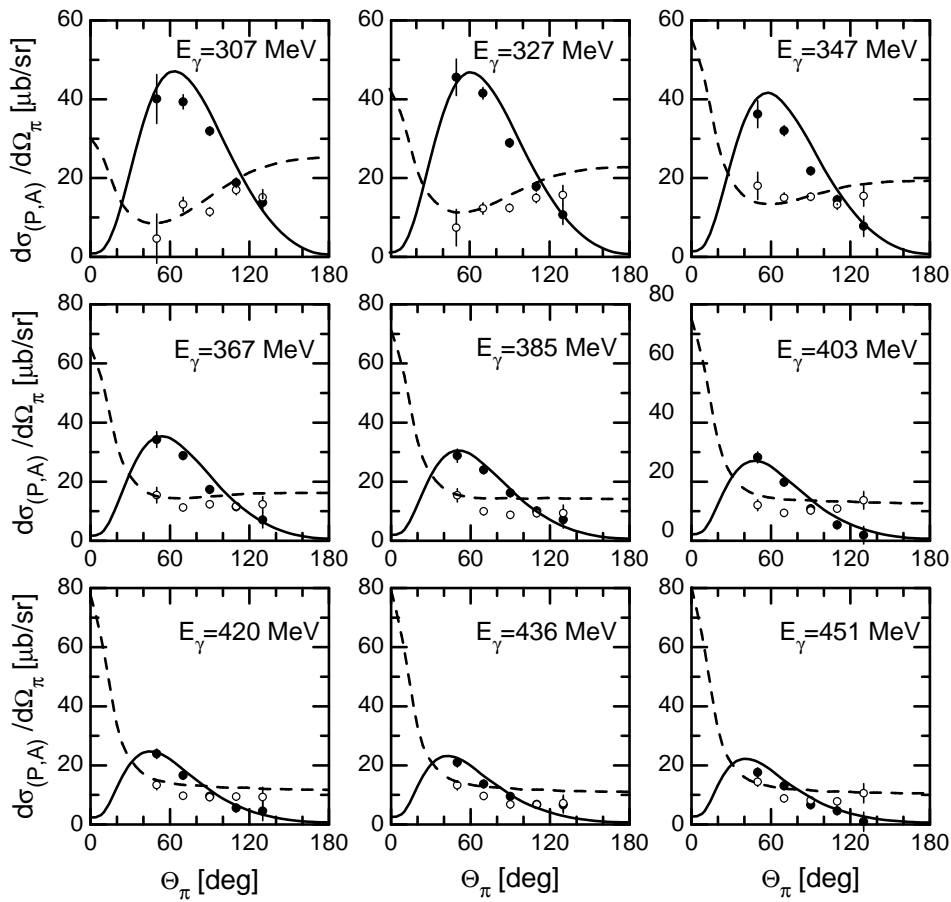}
\caption{
Helicity-dependent differential cross sections
$d\sigma_P /d\Omega_\pi$ and $d\sigma_A /d\Omega_\pi$
for the $\gamma d\to\pi^-pp$ reaction. Solid and dashed curves are
our predictions for $d\sigma_P /d\Omega_\pi$ and $d\sigma_A /d\Omega_\pi$,
respectively, with the amplitudes $A_i$ and the MAID07 analysis.
Data are from Ref.~\cite{ahrens10} for the parallel ($\bullet$)
and antiparallel ($\circ$) components.
Only statistical errors are shown.
 }
\label{spimn_PA}
\end{figure*}

\begin{figure*}[hbt]
\includegraphics[width=0.7\textwidth]{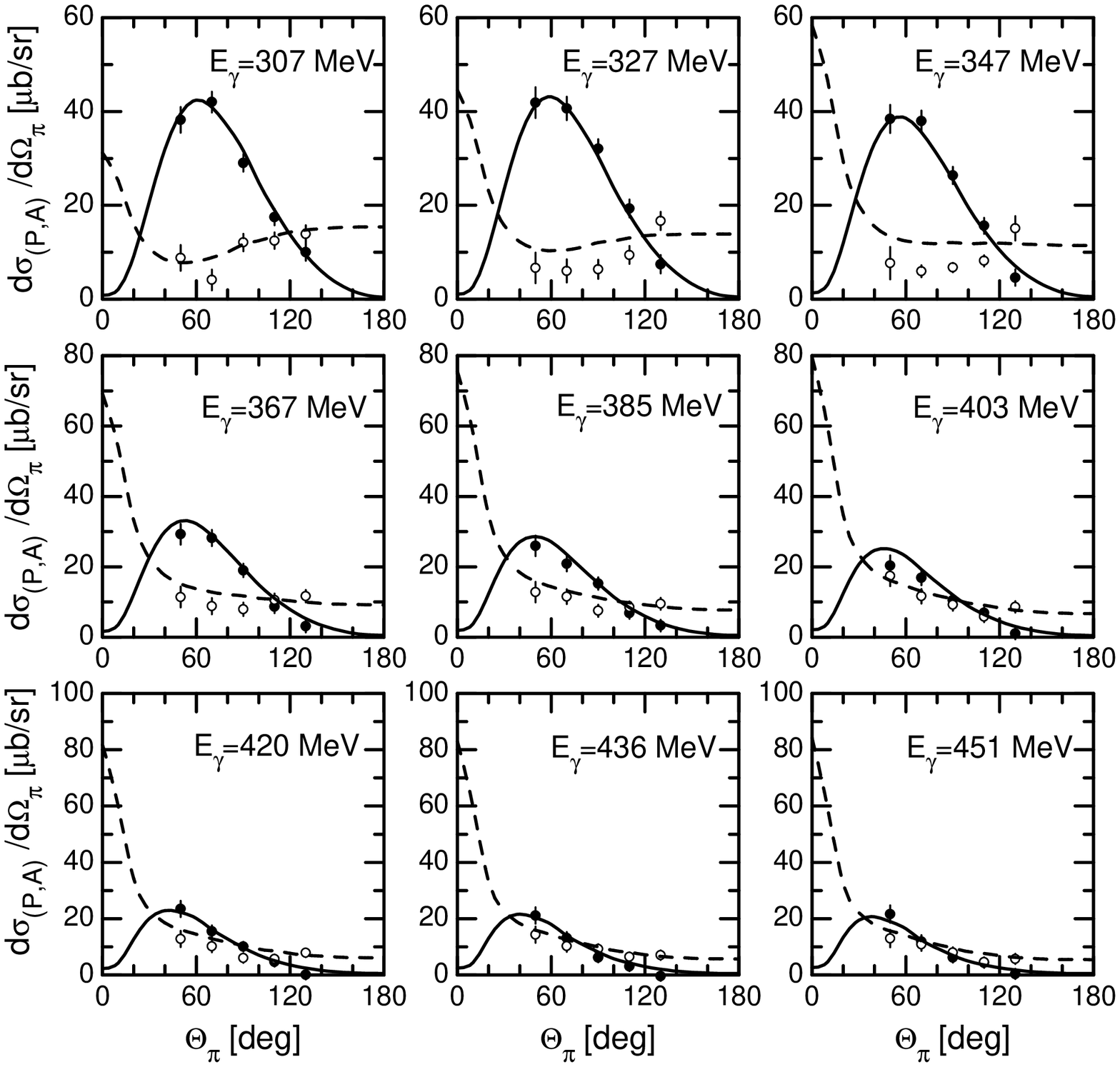}
\caption{
Same as Fig.~\ref{spimn_PA}, but for the $\gamma d\to\pi^+nn$ reaction.
 }
\label{spipl_PA}
\end{figure*}

Figures~\ref{dY_PA}--\ref{spipl_PA} demonstrate our
predictions for the parallel and antiparallel yields and cross sections.
A comparison with the experimental results from Ref.~\cite{ahrens10} is also
presented in these figures.
One can see that the description of the data is quite reasonable,
but not perfect.
Such an agreement could be anticipated for the charged channels,
but it is rather unexpected for the $\pi^0np$ reaction.
Currently, we are not aware of the reasons why the model
that describes reasonably well the $P$ and $A$ components of
the polarized yield, nevertheless, fails in reproducing the unpolarized yield.

We do not discuss here the shape of the angular behavior
of the $P$ and $A$ components in the yield and differential
cross section because the corresponding analysis has already
been performed in Ref.~\cite{ahrens10}. As shown in that work,
this behavior can be interpreted in terms of pion photoproduction
on single nucleons, and it is a consequence of the dominance
of quasi-free reaction mechanisms.

\subsection{The $\gamma d\to\pi NN$ contribution to the GDH integral
for the deuteron}
\label{sub:GDHint}

\begin{figure*}[hbt]
\includegraphics[width=0.75\textwidth]{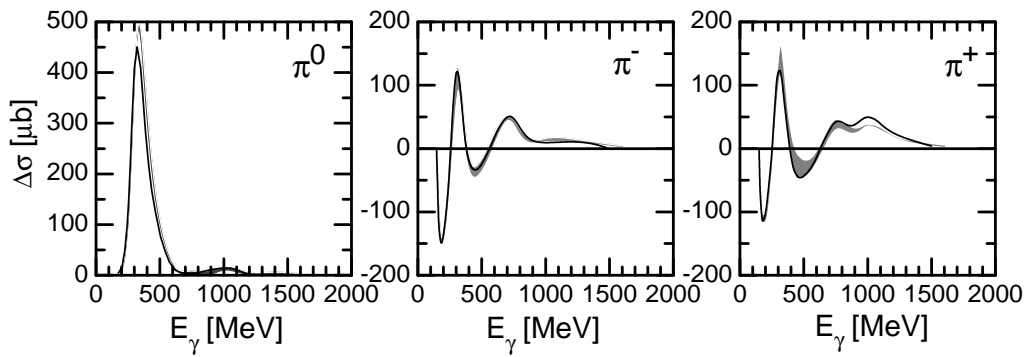}
\caption
{
Difference $\Delta\sigma$
for the separate channels of the $\gamma d \to\pi NN$ reaction.
Shaded areas are our predictions
with different parametrizations of the elementary $\gamma N\to\pi N$
amplitude. Curves are the AFS predictions~\cite{AFS04}.
}
\label{GDHint}
\end{figure*}

Let us discuss now the contribution of the $\gamma d\to\pi NN$
reaction to the deuteron GDH integral and compare our results with
those obtained in Refs.~\cite{{AFS04},darwish07}.
Figure~\ref{GDHint} presents the difference $\Delta\sigma$
as a function of
the photon energy $E_\gamma$ for three channels. As was
shown in Refs.~\cite{AFS04,darwish07}, the behavior
of the difference repeats mainly
that for the elementary reaction $\gamma N\to\pi N$
with some distortion due to nuclear effects. For instance,
the negative value of $\Delta\sigma$ in the charged channels
at threshold energies is because of the dominance of
the $E_{0+}$ multipole producing the antiparallel transition.
The $M_{1+}$ multipole provides the strong positive value
in the $\Delta$ region that Fig.~\ref{GDHint} clearly
demonstrates.
As seen in Fig.~\ref{GDHint}, our results are quite close to those
obtained in Ref.~\cite{AFS04}. Visible difference may be seen only
in the neutral channel in the first resonance region.

The above disagreement also manifests itself in the value of
the GDH integral, integrated up to 1.5 GeV. The upper limit
of 1.5 GeV  is chosen in accordance with an option from Ref.~\cite{AFS04}.
Taking into account the theoretical uncertainties, we obtain the values of
$241\pm 4$, $-20\pm 5$, and $15\pm 20$
(in units of $\mu$b used also below) for the $\pi^0np$, $\pi^-pp$,
and $\pi^+nn$ reactions,
respectively, to be compared to the AFS predictions,
respectively, $222$,
\footnote{In fact, this value has not been given explicitly
in Ref.~\cite{AFS04}. We have extracted it from Fig.~5 of that work
making use of a digitation procedure.},
$-18.94$, and $2.51$. One can see that
our value for the $\pi^0np$ channel exceeds that of AFS by about $10\%$.
Note here that the use of the SAID parametrization allows one to extend
the upper limit in the GDH integral up to 2 GeV. With such an extension
we have not found any visible changes in the result for 1.5 GeV.
This means that the $\pi NN$ term in the deuteron GDH integral essentially
reaches convergence at these energies.

The evaluation of the GDH integral, integrated up to 350 MeV,
has been performed  in Ref.~\cite{darwish07}. Results of that work
with the MAID03 model for the elementary $\gamma N\to\pi N$ operator
are, respectively, $123$, $-29$, and $-9$
for the $\pi^0np$, $\pi^-pp$, and $\pi^+nn$ reactions, which
are in complete agreement with our predictions of
$123\pm 2$, $-34\pm 7$, and $-7\pm 10$, respectively.

We have also studied the influence of the different parametrizations
for the elementary $\gamma N\to\pi N$ amplitude on the value of
the GDH integral, integrated to the maximum energy of 1.65~GeV.
As is seen in Table~\ref{table}, the theoretical uncertainties
in the $\pi^0np$ reaction are quite small. They are more pronounced in
the charged channels, especially in the $\pi^+nn$ one.
This considerable dependence of the results is mainly due to
the multipole analyses,  but the effect of a choice of the
invariant amplitudes is also quite visible.
Note that the noticeable dependence of the nucleon GDH integral
evaluated with
the SAID and MAID analyses was found in Refs.~\cite{SAID,SAID_GDH}.
The sensitivity of the $\gamma d\to \pi NN$
contribution to the deuteron GDH integral, integrated up to 350 MeV,
to a choice of the elementary operator
have been studied in Ref.~\cite{darwish07}, and considerable
dependence of the results obtained with the ELA and MAID03
operators have been found.

\begin{table}
\caption{
Sensitivity of the GDH integral (in $\mu$b), integrated up to 1.65 GeV
in the $\gamma d\to \pi NN$ reaction to the parametrization of
the elementary $\gamma N\to\pi N$ amplitude.
}
\begin{ruledtabular}
\begin{tabular}{lcccc}
Parametrization & $\pi^0np$ & $\pi^-pp$ & $\pi^+nn$  & $\pi NN$ \\
\hline
MAID, $A_i  $    & 235    & --16         & ~6         & 225      \\
MAID, $A_i' $    & 238    & --24         &--4         & 210      \\
SAID, $~A_i $    & 239    & --14         & 35         & 260      \\
SAID, $~A_i'$    & 241    & --22         & 25         & 244
\end{tabular}
\end{ruledtabular}
\label{table}
\end{table}

Keeping in mind all the numbers given in Table~\ref{table},
we can quote our final value as $235\pm 25$~$\mu$b
for the contribution from the $\pi NN$ channel
to the GDH integral for the deuteron, integrated up to 1.65 GeV.
As seen from Table~\ref{table}, the main part of the uncertainties
stems from the $\pi^+nn$ channel.
It is worth mentioning that within the uncertainties of $\pm 25$~$\mu$b,
the AFS result of $27.31$~$\mu$b~\cite{AFS04} for the total contribution
to this integral for the deuteron is compatible with
the GDH sum rule prediction of $0.65$~$\mu$b.

\section{Conclusion}
\label{conclusion}
The central aim of the present work is the study of
the $\gamma d \to\pi NN$ reaction with
the polarized photon and deuteron and its contribution
to the GDH sum rule for the deuteron.
To evaluate the reaction amplitude, we have used
the diagrammatic model elaborated previously in Ref.~\cite{ls06},
which takes into account diagrams corresponding to IA
as well as $NN$ and $\pi N$ interactions in the final state.
For the elementary operator of pion photoproduction on the nucleon,
its on-shell form generated by the recent multipole analyses, SAID and MAID,
has been used. Particular emphasis has been placed on the discussion of
possible uncertainties introduced into the model by this operator.

Comparing our results with the recent experimental data obtained
by the GDH and A2 collaborations in the energy domain
from 300 to 450~MeV~\cite{ahrens06,ahrens09,ahrens10},
we have rediscovered a known problem consisting in the sizable overestimation
of the unpolarized data for the neutral channel. At the same time,
the unpolarized data for the charged channels are reasonably reproduced.

We have found that the helicity-dependent differential
yield difference $\Delta Y/d\Omega_p$ for the neutral channel
and differential cross section difference $\Delta \sigma/d\Omega_\pi$ for
the charged channels are reproduced fairly well.
Also, the model provides a quite satisfactory description of
the separate parallel and antiparallel yields and cross sections
in all the channels.

We have also studied the uncertainties arising from the form of
the elementary operator of pion production on the nucleon.
Regarding the unpolarized observables, the sizable
dependence has been found only for the charged channels
at the lowest energy of about 300~MeV. The difference
$\Delta \sigma/d\Omega_\pi$ manifests the weak sensitivity to
the operator in the $\pi^-pp$ channel in the full energy domain,
and the difference $\Delta Y/d\Omega_p$ in the $\pi^0np$ channel does
the same below 350~MeV. However, a large sensitivity has been found
in the $\pi^+nn$ reaction from 300 to 450~MeV at
$45^\circ <\Theta_\pi <125^\circ$.

Within our model, we have evaluated the contribution
of the $\pi NN$ channel to the GDH integral for the deuteron,
integrated up to 1.65~GeV. In addition, the influence of the
elementary operator on its value has been estimated.
Our final result is $235\pm 25$~$\mu$b.
The uncertainty of $\pm 25$~$\mu$b originates mainly from
the $\pi^+nn$ channel. Accepting these values as
error bars for the value of $27.31$~$\mu$b obtained
in the AFS framework~\cite{AFS04} for the deuteron GDH integral,
one can conclude that within these error bars, the AFS
result is compatible with the GDH sum rule prediction
of $0.65$~$\mu$b.

Further improvements of the present model are needed
to resolve, at last, the problem with the strong overestimation
of the unpolarized cross section in the $\pi^0np$ channel.
In particular, one should evaluate contributions of the two-loop
diagrams which take into account effectively the absorption
of the photoproduced $\pi$ meson by the $NN$ pair.
On the experimental side, it would be very important to continue
polarization measurements in the $\gamma d\to\pi NN$ channel.
Specifically, data on the tensor asymmetry $T_{20}$
together with these on the unpolarized cross section could provide
the currently absent information on the $m_d=0$ component
of the reaction amplitude.

\begin{acknowledgments}

I am very grateful to A. I. Fix and P. Pedroni for valuable
discussions. This work was supported by Belarusian Republican
Foundation for Fundamental Research under Grant No. F09-051.

\end{acknowledgments}

\end{document}